# Calculation of surface tension via area sampling

Jeffrey R. Errington[*] and David A. Kofke[‡]

*Department of Chemical and Biological Engineering,
University at Buffalo, The State University of New York, Buffalo, NY*

## Abstract

We examine the performance of several molecular simulation techniques aimed at evaluation of the surface tension through its thermodynamic definition. For all methods explored, the surface tension is calculated by approximating the change in Helmholtz free energy associated with a change in interfacial area through simulation of a liquid slab at constant particle number, volume, and temperature. The methods explored fall within three general classes: free-energy perturbation, the Bennett acceptance-ratio scheme, and the expanded ensemble technique. Calculations are performed for both the truncated Lennard-Jones and square-well fluids at select temperatures spaced along their respective liquid-vapor saturation lines. Overall, we find that Bennett and expanded ensemble approaches provide the best combination of accuracy and precision. All of the methods, when applied using sufficiently small area perturbation, generate equivalent results for the Lennard-Jones fluid. However, single-stage free-energy-perturbation methods and the closely related test-area technique recently introduced by Gloor et al. [J. Chem. Phys. **123**, 134703 (2005)], generate surface tension values for the square-well fluid that are not consistent with those obtained from the more robust expanded ensemble and Bennett approaches, regardless of the size of the area perturbation. Single-stage perturbation methods fail also for the Lennard-Jones system when applied using large area perturbations. Here an analysis of phase-space overlap produces a quantitative explanation of the observed inaccuracy, and shows that the satisfactory results obtained in these cases from the test-area method arise from a cancellation of errors that cannot be expected in general. We also briefly analyze the variation in method performance with respect to the adjustable parameters inherent to the techniques.

---

[*] Email address: jerring@buffalo.edu
[‡] Email address: kofke@buffalo.edu

# I. Introduction

The thermodynamic and kinetic properties of a liquid-vapor interface play a key role in many scientific phenomena and industrial applications. For example, knowledge of a system's interfacial properties is important when studying the onset of a phase transition through homogeneous nucleation[1] as well as in the design of products such as foaming hand soap and insulating foam sealants. One of the more notable interfacial properties is the surface tension.[2] Although the importance of this quantity has long been established, its determination through molecular simulation remains a challenging task. In this work, we examine the performance of several computational techniques in which the surface tension is evaluated through its thermodynamic definition. These techniques share the common feature of sampling the properties of a system at two or more interfacial areas, and we therefore refer to the methods as "area sampling" techniques.

There are several general approaches that one can employ to evaluate the surface tension. Perhaps the most common route to this property is through the mechanical definition.[3-11] For a planar interface, the surface tension is related to an integral of the difference between the normal and tangential components of the pressure tensor along the direction perpendicular to the interface. The pressure tensor is typically evaluated using the molecular virial.[12] While this method has proven useful for a number of applications, there are a several restrictions that detract from its general applicability. Notably, the pressure tensor can be difficult to evaluate for complex or discontinuous potentials as well as at non-planar interfaces. In addition, the interface can be difficult to establish and maintain at relatively high temperatures along the liquid-vapor saturation line.

A second general method for obtaining surface tension values utilizes the finite-size-scaling formalism of Binder.[13] In this approach, apparent system-size-dependent interfacial tensions are calculated for a range of system sizes, which are subsequently used to extrapolate the true infinite-system-size surface tension through finite-size scaling.[14] Apparent surface tension values are related to the free-energy barrier between vapor and liquid phases, which is extrapolated from a density probability distribution at saturation conditions. The method is particularly appealing at relatively high temperatures where the heterogeneous region between two coexisting phases is comparatively easy to sample for a wide range of system sizes. For temperatures at which the technique is applicable, the approach produces surface tension values with a high degree of precision. Recent fluids for which the method has been used successfully include the Lennard-Jones,[14-16] square-well,[17] *n*-alkane series,[18] and Lennard-Jones mixtures.[19] Weaknesses of the method include (1) the computational cost associated with evaluating apparent free-energy barriers for multiple system sizes and (2) the difficulty in computing density-dependent free energy curves at low temperatures. While recently introduced flat-histogram methods[20-30] help considerably in solving the latter problem,[14] the difficulty associated with sampling the array of heterogeneous structures[31] that one must traverse for large systems at low temperatures can prove prohibitive.

The third general route to the surface tension, and the one we focus on exclusively in this work, stems from the thermodynamic definition of this property.[32-34] When adopting this approach, the surface tension is determined by approximating the change in a system's Helmholtz free energy with a corresponding change in its interfacial surface area, while holding the system at constant mole number, volume, and temperature. In contrast to the mechanical and finite-size scaling routes, this method has seen little use until recently. The appealing aspect of this approach over the mechanical route is that one can determine the surface tension without performing the often-difficult calculation of the pressure tensor. In return, one is presented with the challenge of accurately determining the free-energy difference between two systems separated by a small difference in interfacial area. Note that, similar to the mechanical approach, the thermodynamic route will prove difficult to implement at relatively high temperatures due to challenges associated with stabilization of the liquid slab.

To date, two approaches have been adopted for calculation of the free-energy differences necessary for the thermodynamic approach. Salomons and Mareschal proposed a technique[33] in which this quantity is computed using Bennett's method.[35] Although this method is relatively straightforward to implement, it has not been applied frequently in practice. Gloor et al. recently revisited calculation of the surface tension using the thermodynamic route, and proposed a so-called test-area technique[36] in which the free-energy difference is obtained through free-energy perturbation.[37] Conceptually, the test-area approach is similar to the Widom test-particle method[38] for computation of the chemical potential. In practice, the test-area technique is straightforward to implement, as just a single simulation is needed.

In this work, we investigate the accuracy and precision of several techniques designed to calculate the surface tension through its thermodynamic definition. Free-energy perturbation methods, including the test-particle approach introduced by Gloor et al.,[36] as well as techniques based on Bennett's scheme[33,35] are considered. In addition, we introduce an alternative method, based on the expanded ensemble framework,[39] for evaluating the surface tension. The expanded ensemble approach described here is conceptually similar to that introduced by Lyubartsev et al. for evaluation of the chemical potential.[40] The truncated Lennard-Jones and square-well fluids serve as model systems to test the performance of each method. Selecting these two fluids enables us to directly compare our results with the recent study of Gloor et al.[36] In addition, the finite range of these potentials allows us to complete the simulations without addressing the complications inherent in computing long-range energies in heterogeneous systems.[10,16,41]

The paper is organized as follows. In the next two sections we describe the methods studied in this work and the manner in which our simulations were conducted. We then examine the performance of eight variations of free-energy perturbation, Bennett's method, and the expanded ensemble approaches for computation of the surface tension. The paper concludes with a discussion of the results and several recommendations for computing surface tensions with area-sampling techniques.

# II. Methods

We now describe the techniques examined in this work for calculating the surface tension through molecular simulation. In all cases, relevant quantities are evaluated through canonical ensemble simulations[42] in which the number of particles $N$, temperature $T$, and volume $V$ are held constant. Simulations are conducted in a rectangular parallelepiped cell designed such that the distance in the $z$-direction $L_z$ is approximately a factor of three larger than the box length in the equidistant $x$- and $y$-directions. At the temperatures studied, particles organize into a liquid slab, creating two interfacial regions within the simulation cell, such that the directions normal to the interfaces align with the $z$-axis. The total interfacial area $a$ between the liquid and vapor phases is given by twice the cross-sectional area of the simulation cell $\boldsymbol{a} = 2A = 2L_x L_y$. For all techniques considered, the surface tension $\gamma$ is calculated by estimating the change in the Helmholtz free energy $F$ upon a change in the interfacial area $\boldsymbol{a}$,

$$\gamma = \left( \frac{\partial F}{\partial \boldsymbol{a}} \right)_{N,V,T} \cong \left( \frac{\Delta F}{\Delta \boldsymbol{a}} \right)_{N,V,T}. \qquad (1)$$

The differences in the various techniques examined here stem from the approach used in evaluating the Helmholtz free energy difference $\Delta F$. In what follows, we describe several methods for evaluating this quantity. In most cases, the techniques have been described in detail in previous publications, and therefore we simply highlight the salient features of those methods here.

### A. Free-energy perturbation

Methods in the first group examined are based on a free-energy perturbation formalism developed by Zwanzig.[37] Within this approach, the free energy difference between a reference system (labeled 0) and a target system (labeled 1) is given by,

$$\Delta F = -kT \ln \left\langle \exp \left[ -\left( E_1 - E_0 \right) / kT \right] \right\rangle_0, \qquad (2)$$

where $k$ is Boltzmann's constant, $E_1$ and $E_0$ are instantaneous configurational energies of the reference and target systems, respectively, and $\langle x \rangle_j$ indicates an ensemble average of the quantity $x$ obtained from a simulation that samples system $j$. To determine a surface tension, we require the free energy difference between two systems with different interfacial areas and equivalent particle numbers, volumes, and temperatures. In practice, the free energy is computed by performing a simulation with a reference system with an interfacial area $\boldsymbol{a}_0$, during which periodic perturbations are made to a system with interfacial area $\boldsymbol{a}_0 + \Delta \boldsymbol{a}$, subject to the constraints identified above (to maintain the system volume, as the area is contracted or expanded, the dimension perpendicular to the surface is perturbed in the opposite manner). Although careful consideration has to be given with respect to the value selected for $\Delta \boldsymbol{a}$, typically one has a fair degree of latitude in choosing this value. Below, we use the notations PP and PN to denote calculations in which $\Delta \boldsymbol{a}$ is positive and negative, respectively.

As documented by Gloor et al.,[36] this approach for evaluating surface tension is analogous to the Widom test-particle insertion (or deletion) technique[38,42] for calculating the chemical potential. The advantage of the free-energy perturbation scheme is that one needs to perform only a single simulation (of the reference system) from which all necessary information is obtained. The disadvantage stems from the tendency of free-energy perturbation approaches to yield biased, or inaccurate results.[43-57] The presence of bias is readily identified by considering the free energy differences provided by forward ($0 \rightarrow 1$) and backward ($1 \rightarrow 0$) perturbations. Thermodynamic consistency requires that these results be equal in magnitude and opposite in sign. When the results differ it is possible that both are incorrect, perhaps in roughly equal magnitude (symmetric bias); but it is just as well possible that one is correct and the other is in error (asymmetric bias). Without introducing other measures there is no way to reconcile such a difference.

### B. Test-area method

Gloor et al. recently introduced a so-called test-area (TA) simulation method[36] for evaluating interfacial tensions. Within this scheme, the free-energy differential $\left(\partial F/\partial \boldsymbol{a}\right)_{N,V,T}$ is approximated through a combination of free-energy differences obtained from both forward and reverse perturbations from a specified reference system. To implement this technique, a single simulation is performed in which a reference system 0 samples configurations relevant to a system with interfacial area $\boldsymbol{a}_0$. During the simulation perturbations to a system 1, with a larger interfacial area $\boldsymbol{a}_1 = \boldsymbol{a}_0 + \Delta\boldsymbol{a}$, are used to approximate a free-energy difference $\Delta F_{01}$ using Equation (2). Similarly, perturbations to a system 2, with a smaller interfacial area $\boldsymbol{a}_2 = \boldsymbol{a}_0 - \Delta\boldsymbol{a}$, are used to approximate a free-energy difference $\Delta F_{02}$. The interfacial tension is then approximated by averaging the magnitude of the two free energy differences,

$$\gamma = \frac{1}{2\Delta\boldsymbol{a}}\left(\Delta F_{01} - \Delta F_{02}\right). \tag{3}$$

From a mathematical perspective, the Gloor et al. approach can be thought of as analogous to a central difference approximation for numerical differentiation,[58] whereas a unidirectional free-energy perturbation approach is akin to a forward difference approximation. Numerically, forward and central difference schemes correspond to zeroth- and first-order derivative approximations, respectively.[58] In terms of the chemical potential calculation discussed above, the Gloor et al. approach is analogous to estimating a free-energy difference by averaging the results from test-particle-insertion and test-particle-deletion perturbations. Intuitively, this scheme offers the possibility of acquiring similar accuracy as that provided by the Bennett approach (see below) through the use of just a single simulation (Bennett's method requires two simulations). However, as we will show below, the test-area method provides inaccurate interfacial tension values in cases where asymmetric bias is present in the free-energy calculations.

### C. Bennett's method

The Bennett, or acceptance ratio, method[35,42] is another general technique for determining a free energy difference between two systems (say 0 and 1). In Bennett's scheme, the free energy difference is obtained apparently again by considering perturbations from both system 0 to system 1 ($0 \rightarrow 1$) and from system 1 to system 0 ($1 \rightarrow 0$). The working equation is given by,

$$\Delta F = -kT \ln\left[\frac{\left\langle f\left(E_1 - E_0 - C\right)\right\rangle_0}{\left\langle f\left(E_0 - E_1 + C\right)\right\rangle_1}\right] + C, \tag{4}$$

where the Fermi-Dirac function $f(x) = 1/\left(1 + \exp\left(x/kT\right)\right)$ and $C$ is an adjustable parameter. The optimal value for $C$ is provided by satisfying the following relationship,

$$\sum_{n_{10}} f\left(E_0 - E_1 + C\right) = \sum_{n_{01}} f\left(E_1 - E_0 - C\right), \tag{5}$$

where $n_{10}$ and $n_{01}$ represent the number of perturbations from $1 \rightarrow 0$ and $0 \rightarrow 1$, respectively. Provided the frequency of observing a given energy change $E_1 - E_0$ (and $E_0 - E_1$) as a result of a perturbation is collected during a simulation, the constant $C$ can be obtained during a post-simulation analysis.[42]  In some cases an "unoptimized" Bennett's method is employed in which the constant $C$ is taken to be zero.  We return to a discussion regarding this approach below, drawing a connection between it and a variant of an expanded ensemble technique we examine.

Although Bennett's method has the appearance of calculating perturbation averages involving just the 0 and 1 systems, it is more appropriately viewed as a two-stage free-energy calculation, in which perturbations are performed from the 0 and 1 systems into a common intermediate.[43,44,46,49,53,54,56,57]  The difference $F_1 - F_0$ is thus computed as the sum of differences between each and the intermediate (label it $X$): $(F_1 - F_X) + (F_X - F_0)$.  Bennett's optimization implicitly selects an intermediate system that ensures good results are obtained for both of these perturbations.  Application of the method requires two simulations, sampling the 0 and 1 systems, respectively, with each perturbing into the intermediate system that is defined purely in terms of $E_0$ and $E_1$.

Although others[32] have incorporated the basic principles of the Bennett method within a scheme for calculating the surface tension, Salomons and Mareschal[33] were the first to apply this method within the general framework explored here.  Similar to the free energy perturbation technique discussed above, one typically has a fair degree of latitude in choosing the difference in interfacial area $\Delta a = a_1 - a_0$.  The advantage of Bennett's method is its superior accuracy relative to simple free-energy perturbation.  Studies focused on evaluation of the chemical potential[48,53] have shown that the Bennett technique is capable of producing a correct answer in cases where simple free-energy perturbation fails (e.g. SPC water[59]) due to the asymmetry issues discussed above.  The disadvantage is that two simulations need to be performed in comparison to one for both unidirectional free-energy perturbation and the test-area method.

### D. Expanded ensemble

A fourth method one could use to evaluate the interfacial tension is through an expanded ensemble simulation.[39]  To the best of our knowledge, this approach has not been used previously within the context of calculating surface tensions.  However, one can readily draw an analogy with application of the expanded ensemble to compute chemical potentials.[40,60-63]  The idea behind the expanded ensemble technique is to connect two systems of interest through a series of subensembles, which gradually transform one system of interest into the other, and vice versa.  In the case of the chemical potential, the two systems of interest include one containing a single non-interacting ghost molecule and $N$ interacting molecules and another containing $N + 1$ fully interacting molecules.  These systems are connected through a series of intermediate subensembles in which one of the $N + 1$ molecules partially interacts with the other $N$ molecules.  To compute a surface tension, we connect two systems with different interfacial areas, but with an identical number of particles, volume, and temperature.  This task is accomplished by establishing a sequence of subensembles in which the interfacial area is linearly increased from a minimum value of $a_0$ to a maximum value of $a_M$ in a series of $M$ steps of size $\Delta a$.

Implementation of the expanded ensemble involves conducting a standard Monte Carlo canonical simulation with the additional requirement of trial moves that attempt to shift the system from one subensemble to another.  Within the current context, sampling progresses such that the $i^{\text{th}}$ subensemble is visited with a frequency related to the free energy cost of establishing an interface of area $a_i$.  If we label the subensembles with integer values $m$ that range from $m = 0$ to $m = M$, the free-energy difference between any two subensembles $m$ and $q$ can be expressed in terms of the probabilities $\Pi_m$ and $\Pi_q$ of the system visiting these subensembles,[39]

$$\Delta F = F_m - F_q = -kT \ln\left(\Pi_m / \Pi_q\right). \tag{6}$$

Perhaps the most straightforward means to estimate the probability distribution $\Pi_m$ is through the use of a visited-states approach[21] in which one extracts the necessary information from a histogram $H_m$ of the number of times the Markov chain visits each subensemble during a simulation. Given that the histogram provides a relative probability distribution, the free energy difference between two systems with a difference in interfacial area described by $\boldsymbol{\mathcal{A}}_M - \boldsymbol{\mathcal{A}}_0 = M\Delta\boldsymbol{\mathcal{A}}$ is given by,

$$F_M - F_0 = -kT \ln\left(H_M / H_0\right). \tag{7}$$

The probability distribution can also be approximated using a transition-matrix scheme.[21-25] Within this approach, the relative probabilities of two adjacent subensembles $m$ and $q$ are related to the associated Monte Carlo transition probabilities $P_{m \to q}$ and $P_{q \to m}$ through the detailed balance expression,

$$\frac{\Pi_m}{\Pi_q} = \frac{P_{q \to m}}{P_{m \to q}}. \tag{8}$$

All trial moves that attempt to move the system from subensemble $m$ to $q$ during a Monte Carlo simulation are used to compute the transition probability $P_{m \to q}$,[24,25]

$$P_{m \to q} = \left\langle a_{m \to q}^{\text{Met}} \right\rangle_m, \tag{9}$$

where $a_{m \to q}^{\text{Met}}$ is the conventional Metropolis acceptance probability,[64] which is given by,

$$a_{m \to q}^{\text{Met}} = \min\left\{1, \exp\left[-\left(E_q - E_m\right)/kT\right]\right\}, \tag{10}$$

where $E_m$ and $E_q$ are instantaneous configurational energies of subensembles $m$ and $q$, respectively. The free energy difference between the two extreme subensembles, associated with a difference in interfacial area of $\boldsymbol{\mathcal{A}}_M - \boldsymbol{\mathcal{A}}_0 = M\Delta\boldsymbol{\mathcal{A}}$, is obtained by summing free energy differences between adjacent subensembles,

$$F_M - F_0 = -kT \sum_{m=0}^{M-1} \ln\left[\frac{\left\langle a_{m \to m+1}^{\text{Met}} \right\rangle_m}{\left\langle a_{m+1 \to m}^{\text{Met}} \right\rangle_{m+1}}\right]. \tag{11}$$

Below, we use the notations EV and ET to denote visited-states-based and transition-matrix-based expanded ensemble approaches.

The key operational expression for the expanded ensemble provided by Equation (6) suggests a possible limitation in directly applying this method. As the free-energy difference between any two states increases, the relative frequency with which the higher free energy state is visited decreases exponentially. As a result, long simulation lengths are required to calculate with sufficient precision the likelihood of visiting a subensemble with a relatively high free energy. The probability distribution can often be obtained in a more efficient manner by introducing a weighting scheme that produces a uniform sampling of all subensembles. This aim is achieved by coupling the expanded ensemble technique with a multicanonical sampling procedure.[20]

Implementation of the multicanonical scheme requires each subensemble to be assigned a weight $\eta_m$ that is inversely proportional to the frequency of observing the subensemble during a conventional simulation,

$$\eta_m = -\ln \Pi_m = F_m \big/ kT \, . \tag{12}$$

When utilizing a weighting function one modifies the acceptance probability for Monte Carlo trial moves to account for the weight. For all trial moves that attempt to change the subensemble identity, the multicanonical Metropolis acceptance probability is,

$$a_{m \to q}^{\text{Met, multi}} = \min \left\{ 1, \exp \left[ -\left( E_q - E_m \right) \big/ kT + \eta_q - \eta_m \right] \right\} . \tag{13}$$

In addition, the visited-states expression provided in Equation (7) for the subensemble free energies is modified as follows,

$$F_M - F_0 = -kT \left[ \ln \left( H_M^{\text{multi}} \big/ H_0^{\text{multi}} \right) - \left( \eta_M - \eta_0 \right) \right], \tag{14}$$

where $H_m^{\text{multi}}$ is the visited-states histogram collected during weighted multicanonical sampling. Finally, note that the relevant quantity for calculating subensemble free energies using the transition-matrix approach remains the conventional acceptance probability provided by Equation (10). During a multicanonical transition-matrix simulation the Markov chain evolves according to Equation (13), while the conventional expression is used to accumulate the true transition probabilities $P_{m \to q}$.

The multicanonical sampling scheme requires one to assign weights based on the probability distribution $\Pi_m$, precisely what we are attempting to calculate with the expanded ensemble simulations. Given that $\Pi_m$ is not known at the outset of a simulation, in general, a refinement process must be implemented in which the estimate for the probability distribution, and therefore the weighting function, is periodically improved. For the calculations performed in this work, obtaining a reasonable weighting function did not prove prohibitive. Typically, after a relatively small fraction of the overall CPU time the weighting function converged to provide approximately uniform sampling. The majority of the CPU time was spent refining histogram values or transition probabilities to reduce uncertainties in the total free energy difference. Below we first describe a systematic self-adaptive approach for developing the weighting function and then discuss simpler schemes that can be used to obtain a reasonable estimate of this quantity.

The transition-matrix framework[24,25] described above provides an efficient means to obtain a suitable estimate for the weighting function. A standard expanded ensemble simulation is performed in which subensemble transition probabilities are collected according to Equation (9). At the outset of a simulation, all values of $\eta_m$ are set to zero and the system preferentially visits those subensembles with the lowest free energies. After a number of simulation cycles, accumulated acceptance probabilities are used to estimate the probability distribution and weighting function using Equations (8-12). During the subsequent phase of the simulation, the now non-uniform weighting function encourages the system to visit a broader distribution of subensembles. After several updates of the weighting function, nearly uniform sampling is typically realized. Due to the fact that unbiased acceptance probabilities continue to be used to estimate the probability distribution, one has the ability to periodically update (improve) the weighting func-tion using refined estimates of the transition probabilities without having to discard previously-collected data.

From a practical perspective there are a number of approximate schemes that one could use to generate an initial guess of the weighting function. Recall that for the expanded ensemble approach proposed here the interfacial areas of the subensembles scale linearly with the subensemble identity. It follows that the free energy should vary linearly with subensemble identity. As a result, the only quantity necessary to describe the weighting function is its slope $(\Delta F/kT)/\Delta \boldsymbol{a}$ (the linear-line intercept is arbitrary). A reasonably accurate initial value for the slope can be obtained from an estimate of the surface tension at the temperature of interest. If such a value is obtained, one can establish near-uniform sampling at the outset of a simulation. A reasonable estimate could be obtained from interpolation/extrapolation of surface tension values calculated at other temperatures, from a theoretical approach, or from experimental data when working with real fluids.

### E. Connection between the expanded ensemble and Bennett's method

Before closing our discussion on methods, we explore an interesting connection between Bennett's method and the transition matrix version of the expanded ensemble technique. By way of background, first consider the general expression for the free-energy difference between adjacent subensembles in the expanded ensemble,

$$F_q - F_m = -kT \ln\left[\frac{P_{m \to q}}{P_{q \to m}}\right] = -kT \ln\left[\frac{\langle a_{m \to q}\rangle_m}{\langle a_{q \to m}\rangle_q}\right], \qquad (15)$$

where $a_{m \to q}$ is a generalized Monte Carlo acceptance probability for a trial move that attempts to change the subensemble identity from $m$ to $q$. In the expanded ensemble discussion above we described a procedure that utilizes the Metropolis algorithm[64] for propagation of the Markov chain. However, alternative algorithms can be employed. One such procedure is the Barker sampling technique,[12,65] which is characterized by an acceptance probability based on the Fermi-Dirac function,

$$a_{m \to q}^{\text{Barker}} = f\left(E_q - E_m\right) = \frac{1}{1 + \exp\left[\left(E_q - E_m\right)/kT\right]}. \qquad (16)$$

To make a connection with Bennett's method, we now consider an expanded ensemble limited to two subensembles, labeled 0 and 1, coupled with a transition-matrix Monte Carlo procedure employing Barker sampling. Moreover, we restrict the Markov chain such that attempts to change the subensemble identity are performed but never accepted. To obtain a free-energy difference, two independent simulations are completed; one in which subensemble 0 is sampled and transitions to subensemble 1 are attempted with the restriction noted above, and a second in which the roles of subensembles 0 and 1 are reversed. Using this procedure, the expression for the free-energy difference simplifies to,

$$\Delta F = -kT \ln\left[\frac{\left\langle a_{0\to1}^{\text{Barker}}\right\rangle_0}{\left\langle a_{1\to0}^{\text{Barker}}\right\rangle_1}\right] = -kT \ln\left[\frac{\left\langle f\left(E_1 - E_0\right)\right\rangle_0}{\left\langle f\left(E_0 - E_1\right)\right\rangle_1}\right], \tag{17}$$

which is the same expression provided in Equation (4) with $C = 0$. Therefore, the unoptimized Bennett method is equivalent to a transition-matrix-based two-state expanded ensemble approach employing Barker sampling. A similar connection with Bennett's method was recognized recently by Fenwick and Escobedo,[66] and prior to that Ferrenberg and Swensen[67] showed that optimal combination of histogram data reduces to the optimized Bennett's method when applied to sampling of only two states. Below, we use the notation BO to denote the optimized Bennett method characterized by Equations (4) and (5) above. Additionally, BB and BM are used to describe unoptimized Bennett-like techniques in which Equation (15) is employed to calculate $\Delta F$ with the Barker and Metropolis acceptance probabilities, respectively.

# III. Simulation Details

We use the square-well and truncated Lennard-Jones fluids to investigate performance of the methods discussed above. For the square-well fluid, the energy of interaction $u^{\text{SW}}$ between two particles separated by a distance $r$ is given by,

$$u^{\text{SW}}\left(r\right) = \begin{cases} \infty & r < \sigma \\ -\varepsilon & \sigma < r < \lambda\sigma \\ 0 & r > \lambda\sigma \end{cases}. \tag{18}$$

For the truncated Lennard-Jones fluid, the energy of interaction $u^{\text{LJ}}$ is given by,

$$u^{\text{LJ}}\left(r\right) = \begin{cases} 4\varepsilon\left[\left(\dfrac{\sigma}{r}\right)^{12} - \left(\dfrac{\sigma}{r}\right)^{6}\right] & r < r_c \\ 0 & r > r_c \end{cases}. \tag{19}$$

In both cases $\varepsilon$, $\sigma$, and $\lambda$ (or $r_c$) set the characteristic energy scale, length scale, and range of interaction, respectively. In this work, we restrict our attention to a square-well fluid for which $\lambda = 1.25$ and a Lennard-Jones fluid with $r_c = 2.5\sigma$. From this point forward, all quantities are made dimensionless using $\varepsilon$ and $\sigma$ as characteristic energy and length scales, respectively. For example, temperature and surface tension are reduced by $\varepsilon/k$ and $\varepsilon/\sigma^2$, respectively.

Five sets of calculations were completed to assess the methods. The basic structure of each algorithm is shown schematically in Figure 1. For each surface tension estimate, a total of $N_{\text{MC}}$ Monte Carlo cycles were completed with interfacial area changes performed on average twice per cycle (a cycle is defined as $N$ Monte Carlo moves). The first and second sets of calculations were employed to evaluate surface tensions using the PP and PN methods described above, respectively. In both cases, a system with interfacial area $a_0$ was sampled for $N_{\text{MC}}$ cycles and perturbations to a system with interfacial area $a_1 = a_0 \pm \Delta a$ were performed. A third set of calculations was used to investigate the TA method. A single simulation was performed in which a system with interfacial area $a_0$ was sampled for $N_{\text{MC}}$ cycles and with equal probability perturbations to a system with an interfacial area of either $a_1$ or $a_2$ were performed on average twice per cycle. The fourth set of calculations was used to examine Bennett's method. Two independent simulations were performed for each estimate of the surface tension. In the first (second) simulation a system with interfacial area $a_0$ ($a_1$) was sampled for $N_{MC}/2$ cycles and perturbations to a system with interfacial area $a_1$ ($a_0$) were performed. During each simulation run both Metropolis and Barker transition probabilities were monitored and the distribution of energy changes resulting from a perturbation was collected. These data were used to generate surface tension values based on the BB, BM, and BO methods. The final set of calculations was designed to evaluate the expanded ensemble method. Simulations were run for a total of $N_{MC}$ cycles with subensemble identity changes attempted on average twice per cycle. As noted above, preliminary runs indicated that relatively little simulation time was required to obtain a reasonable estimate of the weighting function. As a result, at the outset of the simulation we simply specified the weighting function as a linear function with a slope consistent with the value of $\Delta F/\Delta a$ obtained from the optimized Bennett approach. Several combinations of the total number of subensembles $M$ and difference in interfacial area between adjacent subensembles $\Delta a$ were examined.

Simulations were performed in a tetragonal cell with periodic boundary conditions in all directions. In all cases, system 0 was designed such that the length of the cell in the direction perpendicular to the interface was a factor of three larger than the length of the cell parallel to the interface. This ratio was altered as needed to maintain a constant cell volume for all other systems of interest. The Monte Carlo move mix consisted of interfacial area changes (on average two per cycle) and four single-particle displacement move styles. Particle displacement moves were distributed as follows: 50% small-scale displacements in the $x$-$y$ plane of magnitude selected uniformly within the interval $\left[ -\Delta_{xy}, +\Delta_{xy} \right]$, 25% small-scale displacements in the $z$-direction of magnitude selected uniformly within the interval $\left[ -\Delta_z, +\Delta_z \right]$, 10% large-scale displacements in the $z$-direction of magnitude selected uniformly within the interval $\left[ -L_z/2, +L_z/2 \right]$, and 15% AVBMC2 aggregation-volume-bias Monte Carlo moves, as described by Chen and Siepmann,[68,69] to facilitate equilibration and sampling of the heterogeneous systems. The magnitudes of $\Delta_{xy}$ and $\Delta_z$ were adjusted to generate an average acceptance probability of 50%.

For the Lennard-Jones fluid, we used $N = 1372$ particles and set the interfacial area $a_0 = 288$ ($A = L_x L_y = 144$). The simulation length was set to $N_{MC} = 2.5 \times 10^5$ cycles. For the square-well fluid, the parameters were specified as follows: $N = 864$, $a_0 = 200$ ($A = L_x L_y = 100$), and $N_{MC} = 2.5 \times 10^7$. Statistical uncertainties $\sigma_x$ for property $x$ were determined by performing four independent sets of simulations, for a total effort of $4N_{MC}$ cycles at each state point. The standard deviation of the results from the four simulation sets was taken as an estimate of the statistical uncertainty. These simulation details are similar to those utilized by Gloor et al. in a recent study[36] involving the two fluids examined here. One exception of note is that we ran our square-well simulations for a factor of ten longer. This action was taken to reduce statistical uncertainties such that relevant differences between the methods could be clearly identified.

# IV. Results

Table 1 contains surface tension values for the truncated Lennard-Jones fluid at select temperatures generated from calculations based on the eight methods discussed above. This information is displayed graphically in Figure 2, with each estimate normalized by the temperature-specific average surface tension obtained through the eight techniques explored. An interfacial area change of $\Delta a = 0.2$ was used in all cases, and a value of $M = 10$ was selected for the expanded ensemble simulations. The magnitude of the area change is close to that used in the recent work of Gloor et al. for the Lennard-Jones fluid.[36] Our results indicate that all of the methods produce equivalent values for the surface tension with similar uncertainties. To provide a feel for the precision one obtains with the methods studied here, we have placed a characteristic uncertainty bar (the magnitude was taken as the median uncertainty of the eight techniques) to the right of each set of data in Figure 2. For the Lennard-Jones fluid, uncertainties of two percent were commonly obtained for calculations at low to moderate reduced temperatures, with the relative uncertainty increasing as one approached the critical point.

Table 2 and Figure 3 contain surface tension values for the square-well fluid at three temperatures. We used an area change of $\Delta a = 0.02$, again consistent with the study of Gloor et al.,[36] and $M = 10$ for the expanded ensemble simulations. Our calculations reveal clear differences between the techniques. The expanded ensemble and Bennett-based methods produce equivalent estimates for the surface tension, whereas the perturbation-based PP, PN, and TA values are statistically different. The PP and PN estimates (see the inset to Figure 3) are a factor of 20 to 60 larger in magnitude than the others, with the PN technique generating a negative surface tension. The test-area estimates, which stem from an averaging of PP and PN values, are on average 25% larger than the corresponding expanded ensemble results. Finally, we note that our TA results are in agreement with values obtained by Gloor et al.

When performing calculations with the liquid-slab structure employed in both the mechanical and thermodynamic route to the surface tension one is often concerned with the extent to which the interfacial area and/or aspect ratio of the simulation cell influence the result. The expanded ensemble approach provides a means to address this question in a straightforward manner. The expanded ensemble can be used to obtain the system free energy over a relatively wide range of interfacial areas, and the local differential of the free energy with respect to the interfacial area provides an estimate of the surface tension as a function of interfacial area. We provide an example of the results obtained from this type of analysis for the Lennard-Jones system in Figure 4. The data displayed correspond to a system at $T = 0.90$ with $N = 1372$ and $V = 5184$. The surface tension shows an oscillatory behavior for small interfacial areas (large aspect ratios $L_z/L_x$) before reaching a nearly constant value beyond $L_x \approx 10$. Recall that the surface tension values reported in Table 1 were generated with $L_x = 12$. We observed similar oscillatory behavior at large aspect ratios for a system with $N = 500$ and $V = 2187$.

A key adjustable parameter for all of the methods explored here is the magnitude of the interfacial area change $\Delta a$. Figure 5 provides estimates of the surface tension and associated uncertainties as a function of $\Delta a$ for the Lennard-Jones fluid at $T = 0.90$. The perturbation-based methods produce equivalent results up to $\Delta a \approx 4$, with the uncertainties steadily increasing after $\Delta a \approx 2$. The Bennett-based methods retain good accuracy and precision to $\Delta a \approx 6$. Additional calculations show that both the perturbation- and Bennett-based methods produce an accurate surface tension with an uncertainty of 2-4% when employing an area change as low as $\Delta \boldsymbol{a} \approx$ 0.01. The expanded ensemble simulations were again constructed with $M = 10$. Calculations were performed only to $\Delta \boldsymbol{a} = 4$ to keep the interface from becoming too thin. Accurate results are obtained in all cases, and for $\Delta \boldsymbol{a} \geq 0.4$ uncertainties are generally equal to or lower than those of the other methods. Using precision as a basis, expanded ensemble simulations with $\Delta \boldsymbol{a} \leq 0.2$ generally do not perform as well as Bennett-based methods.

Figure 6 contains estimates of the surface tension and associated uncertainties as a function of $\Delta \boldsymbol{a}$ for the square-well fluid at $T = 0.65$. The data indicate that the TA approach returns surface tension estimates that are consistently larger than the expanded ensemble and Bennett-based values, regardless of the interfacial area change adopted. The best combination of precision and accuracy is obtained by using either an expanded ensemble or Bennett-based approach with $\Delta \boldsymbol{a}$ set between 0.02 and 0.2. While the PP and PN methods provide rather precise estimates for the surface tension with an appropriate choice for $\Delta \boldsymbol{a}$, the accuracy of these techniques is very poor. This example underscores one of the unsettling aspects of perturbation-based approaches – these methods can produce seemingly correct estimates with excellent precision that are tremendously inaccurate.

Data generated for the Lennard-Jones and square-well fluids suggest that the Metropolis acceptance probability could be a useful guide in selecting the magnitude of the interfacial area change. For all methods examined, the uncertainties appear to be at a minimum when $a^{\mathrm{Met}}$ is between roughly 20 and 80%. The data also suggest that the lower bound for this range can be extended to 1% when employing expanded ensemble and Bennett-based techniques. Selecting a relatively small value of $\Delta \boldsymbol{a}$ can lead to poor performance. As $\Delta \boldsymbol{a}$ is decreased towards rather small values, progressively more simulation time is required to calculate transition probabilities and Boltzmann factors with the precision required to produce meaningful surface tension estimates. In the case of the expanded ensemble it becomes difficult to distinguish the true probability distribution from the noise associated with a near-random walk along the Markov chain.

We also looked briefly at the relative performance of the expanded ensemble approach with respect to the number of subensembles employed. A series of calculations were completed with the Lennard-Jones system in which the area change was set to $\Delta \boldsymbol{a} = 1$ and the total number of subensembles spanned from $M = 4$ to 46. We did not observe a trend in the performance (based on the statistical uncertainty of the surface tension) of the expanded ensemble method over this range of $M$ values. Overall, the efficacy of the method appeared to be more sensitive to the choice of $\Delta \boldsymbol{a}$ than of $M$.

# V. Discussion

All of the approaches examined in this work are in principle rigorously correct. Provided the relevant details of a Monte Carlo simulation are properly accounted for, one can consider surface tension estimates generated from any of the methods to be exact. In practice however we see that systematic, reproducible bias in results from otherwise correctly implemented simulations can lead to the failure of some approaches. Perturbation-based techniques are particularly bad in this regard. Kofke and coworkers have demonstrated that single-stage free-energy-perturbation calculations provide accurate results only when the phase spaces sampled by the reference and target systems satisfy a subset relationship.[43-57] Specifically, the region of configu-ration space sampled by the target system must be subsumed by that sampled by the reference system. This requirement implies that "forward" and "reverse" perturbation estimates will provide equivalent results only when the configuration spaces relevant to the reference and target systems are nearly coincident. If this condition is not met, at least one of the estimates will be inaccurate. Moreover, there is no fundamental basis to justify an exact (or approximate) canceling of errors when combining forward and reverse perturbation estimates, as is attempted in the test-area approach. An inaccurate result from one or both perturbation directions will generally lead to an inaccurate combined result.

Kofke and coworkers have also used phase space relationships to characterize Bennett's method.[43,44,46,49,53,54,56,57] They have shown that the Bennett technique provides accurate free-energy differences when the configuration spaces of the two systems involved in a calculation simply overlap. This less-restrictive requirement implies that the two systems need only sample a common region of configuration space to produce an accurate result. From an implementation perspective, Bennett's method requires slightly more overhead than free-energy perturbation in that two independent simulations are required instead of one. However, with this minor inconvenience comes a softer requirement for the phase-space relationship between the reference and target systems. Also, note that a molecular dynamics protocol can be used to sample the two systems involved in a Bennett calculation.

Expanded ensemble methods are not known to suffer from a tendency to yield biased results, but they too are limited to application to systems having at least some phase-space overlap. This is certainly the case if the method is applied using only two subensembles, which is the situation most comparable to the Bennett approach. More typically multiple subensembles are used, which makes the method comparable to an equivalently multistaged free-energy perturbation approach, which is not considered in this work. Regardless, if there is not phase-space overlap between any two subensembles there will be little or no transition between them, and the necessary degree of sampling is not achieved, yielding inaccurate results. The imposition of weights is not by itself sufficient to eliminate such problems; instead it is necessary to introduce or redefine subensembles in a way that ensures that adjacent ones exhibit phase-space overlap. The appropriate scheme for weighting the subensembles is not always obvious,[43] and may depend on the overlap features of the subensembles. This is a topic that has not been well studied.

We can now use phase space relationships to interpret the surface tension results generated here. In the case of the Lennard-Jones fluid, all methods generated statistically equivalent surface tensions for small enough values of $\Delta a$. The simultaneous success of both the PP and PN methods suggest that the configuration spaces sampled by two Lennard-Jones systems separated by a small difference in interfacial area $\Delta a$ are nearly coincident. For larger $\Delta a$ the overlap is less satisfactory, and the consequence is an increasing bias in the computed surface tension. The overlap between the phase spaces can be quantified by their relative entropies $s_F$ and $s_R$, which can be evaluated from the dissipated work[52-55]

$$s_F = \frac{1}{kT}\Big[\big\langle E_1 - E_0 \big\rangle_0 - \big(F_1 - F_0\big)\Big]$$

$$s_R = \frac{1}{kT}\Big[\big\langle E_0 - E_1 \big\rangle_1 - \big(F_0 - F_1\big)\Big]$$

(20)

(relative entropies are implicitly made dimensionless by Boltzmann's constant here and in what follows). The quantity $\big\langle E_1 - E_0 \big\rangle_0$ corresponds to the average energy difference between the ref-erence (0) and target (1) systems for forward (PP) perturbations and $\left(F_1 - F_0\right)$ represents the true (unbiased) free energy difference between systems 0 and 1. Analogous definitions hold for the reverse (PN) perturbation. Wu and Kofke developed a semi-empirical bias model that predicts from the relative entropies the amount of bias expected in the free energy for a given amount of sampling.[50-53,55] The model suggests that the bias $B_F$ to be expected from a single-stage free-energy perturbation using $M_p$ samples in the forward direction is given by

$$B_F\left(M_p\right) = -kT\ln\left[\frac{1}{2}\mathrm{erfc}\left(s_F^{1/2}\left[1-\left\{\frac{1}{2s_R}\mathbf{W_L}\left[\frac{1}{2\pi}\left(M_p-1\right)^2\right]\right\}^{1/2}\right]\right)\right] \qquad (21)$$

with $\mathbf{W_L}(x)$ the Lambert W-function, which for large $x$ grows as $\ln(x)$. An analogous formula for the expected bias from perturbation in the reverse direction is obtained by exchanging the "$F$" and "$R$" subscripts. In previous applications we have observed the magnitude of the bias to lie within a narrow band bounded from above by the value obtained from Equation (21). Note that the Lambert function which contains the dependence on $M_p$ is divided by $s_R$, which means that the forward-direction bias is affected more by this reverse-direction relative entropy; likewise the "forward" relative entropy $s_F$ is more significant in determining the reverse-direction bias.

To demonstrate the connection between the bias and the relative entropies, and hence the phase-space overlap, we calculated for the Lennard-Jones model values of $s_F$ and $s_R$ as a function of the area perturbation $\Delta\boldsymbol{a}$. Calculated values range from 0.080 to 30.5 for $\Delta\boldsymbol{a}$ = 0.4 to 8.0 and for a given $\Delta\boldsymbol{a}$ $s_F$ and $s_R$ are not very different from each other (the largest observed difference is 0.4 for $\Delta\boldsymbol{a}$ = 8.0). Using these results in Eq. (21) we can compute the expected bias in the free energy difference for the amount of sampling conducted in our simulations, and from this we can get the expected bias in the surface tension. Results for the surface-tension bias are presented in Figure 7, where we see that the bias model provides a very good characterization of the systematic error in the surface tension. In this case the expected bias for the forward and reverse perturbations are very nearly equal in magnitude (and opposite in sign), which is a direct consequence of the rough equality of $s_F$ and $s_R$.

In the case of the square-well fluid, only the Bennett-based and expanded ensemble techniques produced accurate results. Based on the inaccurate values generated by both the PP and PN techniques, the two configuration spaces sampled by square-well systems separated by a small difference in interfacial area $\Delta\boldsymbol{a}$ do not exhibit a subset relationship, i.e., neither configuration space subsumes the other. The relative entropies $s_F$ and $s_R$ for these systems are both infinite for all $\Delta\boldsymbol{a}$, which means that some bias will be present no matter how much sampling is conducted. This does not imply that the area-perturbed systems have no phase-space overlap with the unperturbed system, but that the non-overlapping regions are, because of the hardness of the potential, absolutely forbidden to one or the other system (indeed it is the existence of overlapping phase-space regions that enables Bennett's method to succeed). Unlike the application to the Lennard-Jones system, the bias is not symmetric, which is demonstrated clearly by the continued presence of bias in the TA result, which averages of the forward and reverse data; the bias remains even for small $\Delta\boldsymbol{a}$. Unfortunately in this case the (infinite) relative entropies cannot provide any clues about the relative magnitudes of the forward and reverse bias. This particular limitation of the relative entropies as overlap measures might be addressed in future refinements of the bias-detection model.

It is interesting to note that different phase space relationships hold for chemical potential and surface tension calculations for the Lennard-Jones fluid. In the former case, test-particle insertion provides an accurate estimate of the chemical potential, whereas test-particle deletion does not.[43,44] This combination suggests that (for densities not too close to freezing) the configuration space sampled by a system with $N + 1$ Lennard-Jones particles is a subset of that sampled by $N$ particles and a single ghost particle. In contrast, the phase spaces involved in the surface tension calculation are nearly coincident for relatively small values of $\Delta a$. These examples point to a broader issue with free-energy perturbation techniques. In general, it is not possible to predict reliably this relationship *a priori*, and thereby secure the accuracy of a calculation before it is performed. In light of this feature of perturbation techniques, we argue that these techniques should always be applied in conjunction with the use of a bias-detection diagnostic. Otherwise one should employ methods that require softer criteria to ensure their validity.

One can readily identify analogies to evaluation of the surface tension through the thermodynamic route. One example involves evaluation of the pressure through consideration of the change in the Helmholtz free energy with volume at constant particle number and temperature. Free-energy perturbation approaches have been proposed for calculating the pressure in this manner.[70-72] For example, De Miguel and Jackson recently used a perturbation approach to calculate the pressure tensor for a hard sphere fluid.[72] We note that approaches analogous to the Bennett method and expanded ensemble technique explored here for calculation of the surface tension could be used to evaluate the pressure through the thermodynamic route as well. Based on the inherent accuracy of these methods in comparison to single-stage perturbation techniques, they are likely to provide a safer means to evaluate the pressure at no additional computational cost.

# VI. Conclusions

We have examined the performance of several techniques aimed at calculation of the surface tension through its thermodynamic definition. We considered methods based on free-energy perturbation, the Bennett acceptance-ratio scheme, and an expanded ensemble approach introduced by us. Surface tension values were generated for both the Lennard-Jones and square-well fluids at three temperatures broadly spaced over their respective liquid-vapor saturation lines. At a single temperature for each fluid we examined how the accuracy and precision of surface tension estimates evolve with the magnitude of the interfacial area change $\Delta a$ for each of the methods explored in this work. Finally, we explored the accuracy of the perturbation- and Bennett-based techniques in terms of the relationship between the phase spaces sampled by the two systems involved in each calculation.

Our study points to a few issues that should be considered when employing area sampling techniques to evaluate a surface tension. First, the perturbation-based PP, PN, and TA methods should be used with caution. Application of these methods with the square-well fluid clearly shows that precise, yet inaccurate, results can easily be obtained with these approaches. In contrast, the expanded ensemble and Bennett techniques provided accurate results with similar precision for both fluids examined here. Second, our analysis of the influence of the magnitude of the interfacial area change $\Delta a$ suggests that one can use the Metropolis acceptance probability as a guide to sizing this parameter.  Specifically, the data indicate that adjustment of $\Delta a$ such that $a^{\text{Met}} \approx 40\%$ leads to optimal precision for the expanded ensemble and Bennett-based techniques.

The thermodynamic-based methods examined here appear to provide a promising alternative to the mechanical-route or Binder approach to evaluation of surface tensions.  These area-sampling methods are relatively straightforward to implement and provide reasonable precision for the simulation time expended.  Moreover, the techniques can be naturally incorporated with a molecular dynamics protocol.  Future studies of interest include application of area sampling methods to more complex fluids, such as mixtures and aqueous solutions, adaptation of area-sampling methods to calculation other interfacial tensions of interest, and a systematic comparison of the merits of the thermodynamic, mechanical, and Binder approaches to evaluation of the surface tension.

# Acknowledgements

We acknowledge the financial support of the National Science Foundation, Grant Nos. CTS-028772 [J.R.E.] and CHE-0626305 [D.A.K.], and J.R.E. further acknowledges support from the Donors of the American Chemical Society Petroleum Research Fund Grant No. 43452-AC5.  Computational resources were provided in part by the University at Buffalo Center for Computational Research.

# Tables

**Table 1.** Surface tension for the truncated Lennard-Jones fluid.

| Method | Surface Tension | | |
|---|---|---|---|
| | $T = 0.70$ | $T = 0.90$ | $T = 1.10$ |
| PP | 0.789(16) | 0.411(8) | 0.092(3) |
| PN | 0.788(15) | 0.410(13) | 0.093(3) |
| TA | 0.791(15) | 0.418(11) | 0.097(5) |
| BB | 0.791(10) | 0.409(9) | 0.094(4) |
| BM | 0.795(13) | 0.412(5) | 0.095(5) |
| BO | 0.790(10) | 0.409(9) | 0.094(4) |
| ET | 0.790(14) | 0.420(3) | 0.097(13) |
| EV | 0.789(17) | 0.418(5) | 0.098(12) |

**Table 2.** Surface tension for the square-well fluid.

| Method | Surface Tension | | |
|---|---|---|---|
| | $T = 0.61$ | $T = 0.65$ | $T = 0.70$ |
| PP | 8.107(25) | 7.454(20) | 6.178(26) |
| PN | -7.361(12) | -6.921(11) | -5.920(14) |
| TA | 0.373(13) | 0.269(16) | 0.128(18) |
| BB | 0.303(18) | 0.219(17) | 0.105(22) |
| BM | 0.303(21) | 0.219(17) | 0.102(23) |
| BO | 0.303(18) | 0.219(17) | 0.105(22) |
| ET | 0.313(16) | 0.204(17) | 0.096(17) |
| EV | 0.314(16) | 0.204(16) | 0.098(16) |

# Figure Captions

**Figure 1.** Schematic description of the eight methods examined in this work. Each box represents a simulation cell. The interfacial area increases from left to right. Arrows with broken lines represent interfacial area changes that are performed but never accepted. Arrows with solid lines represent interfacial area changes that are accepted with the Metropolis probability.[64]

**Figure 2.** Surface tension of the Lennard-Jones fluid as a function of temperature. Each estimate is normalized by the temperature-specific average surface tension from the various methods examined. The uncertainty bars provide a feel for the precision obtained from area-sampling techniques, with the size of the bar taken as the median uncertainty value from the eight methods examined. Each symbol corresponds to the results from a given method, with the method identity provided in the legend. See the text for a description of each method.

**Figure 3.** Surface tension of the square-well fluid as a function of temperature. PP and PN results are not visible within the main panel. Uncertainty bars are shown only for TA and EV results. The inset shows a broader range of values along the ordinate. Symbols are the same as in Figure 2.

**Figure 4.** Surface tension of the Lennard-Jones fluid at $T = 0.90$ as a function of the simulation cell length parallel to the interface. Circles, squares, and diamonds represent data from expanded ensemble simulations restricted to sample cross-sectional areas $A = L_x^2$ within the ranges 36-62, 60-124, and 120-196, respectively.

**Figure 5.** Surface tension (top panel), uncertainty in surface tension (middle panel) and Metropolis acceptance probability collected during the Bennett simulations (bottom panel) as a function of interfacial area change for the Lennard-Jones fluid.

**Figure 6.** Surface tension (top panel), uncertainty in surface tension (middle panel) and Metropolis acceptance probability collected during the Bennett simulations (bottom panel) as a function of interfacial area change for the square-well fluid. PP and PN results are not visible within the top panel. Uncertainty bars are shown only for TA and EV results within the top panel.

**Figure 7.** Absolute value of the bias in the surface tension of the Lennard-Jones fluid computed by single-stage free-energy perturbation, as a function of the area perturbation $\Delta \mathcal{A}$. Symbols indicate the bias observed by averaging results from four independent Monte Carlo simulations for the forward (PP) and reverse (PN) perturbation directions. Lines show the bias expected according to Eq. (21) with $M_p = 500,000$ and using independently calculated values of the relative entropies (lines for PN and PP are indistinguishable on this scale).

# Figure 1

| Method | $\longrightarrow$ Interfacial Area $\longrightarrow$ |
|--------|:----:|
| PP | [ 0 ] ·····$\rightarrow$ [ 1 ] |
| PN | [ 1 ] $\leftarrow$····· [ 0 ] |
| TA | [ 2 ] $\leftarrow$····· [ 0 ] ·····$\rightarrow$ [ 1 ] |
| BB/BM/BO | [ 0 ] ·····$\rightarrow$ $\leftarrow$····· [ 1 ] |
| ET/EV | [ 0 ] $\rightarrow$ $\leftarrow$ [ 1 ] $\rightarrow$ $\leftarrow$ [ 2 ] $\rightarrow$ $\leftarrow$ ● ● ● $\rightarrow$ $\leftarrow$ [ *M* ] |

## Figure 2

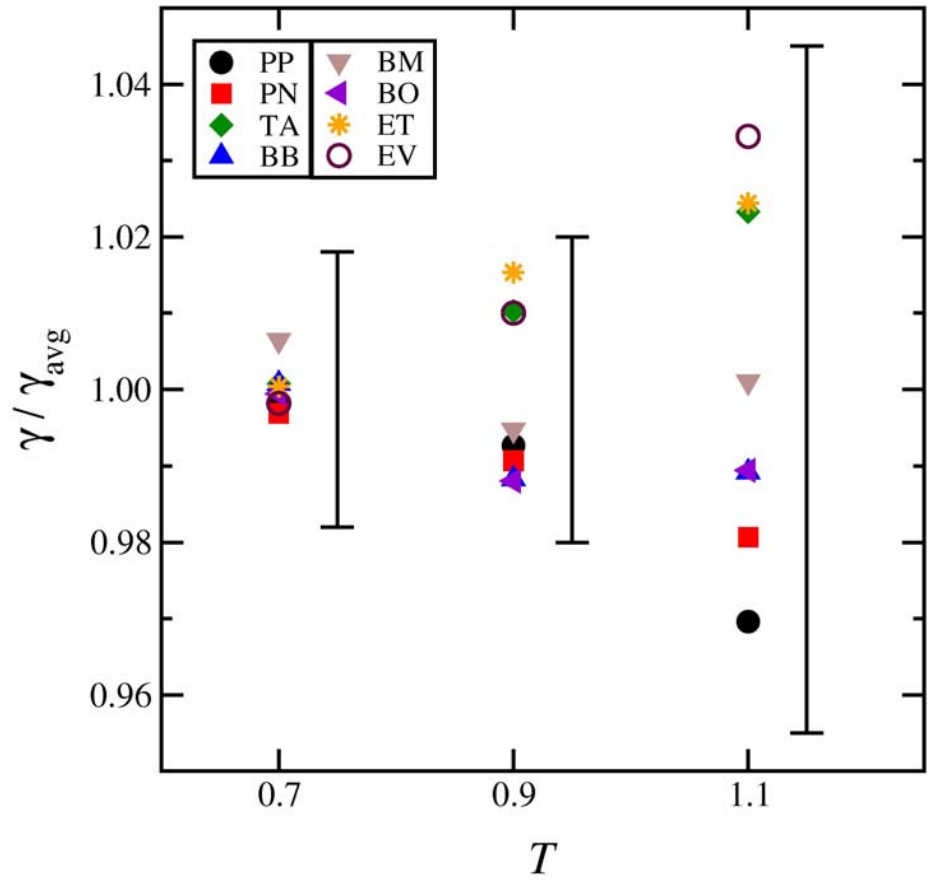

## Figure 3

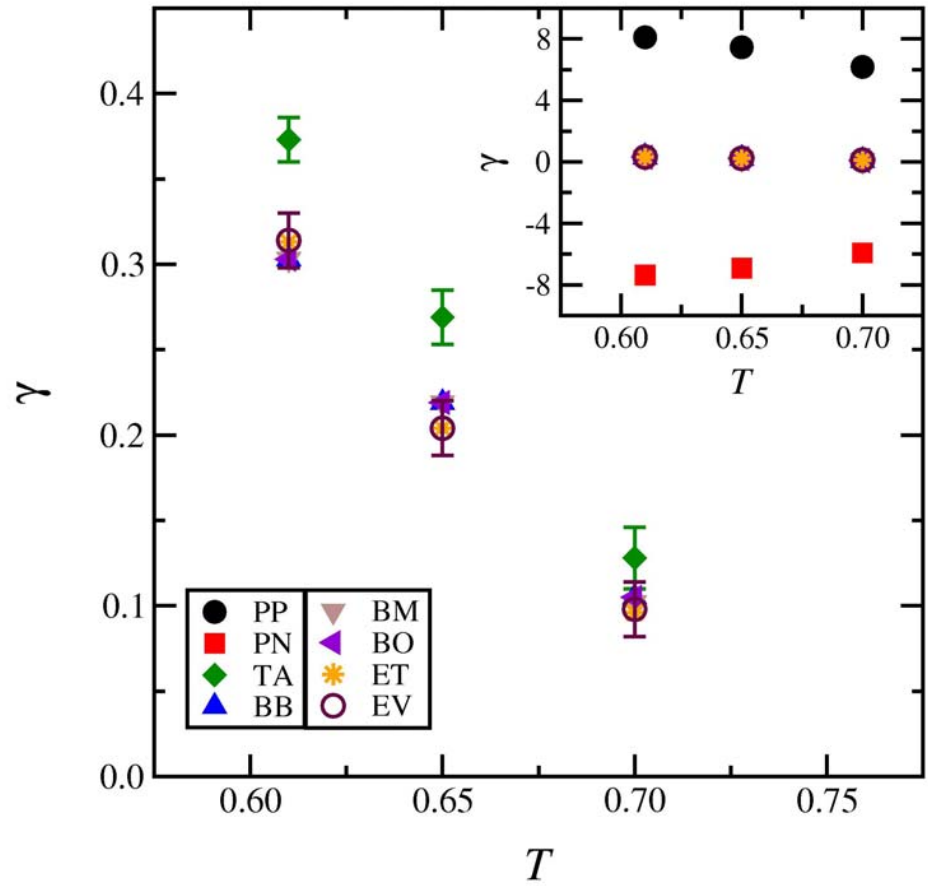

Figure 4

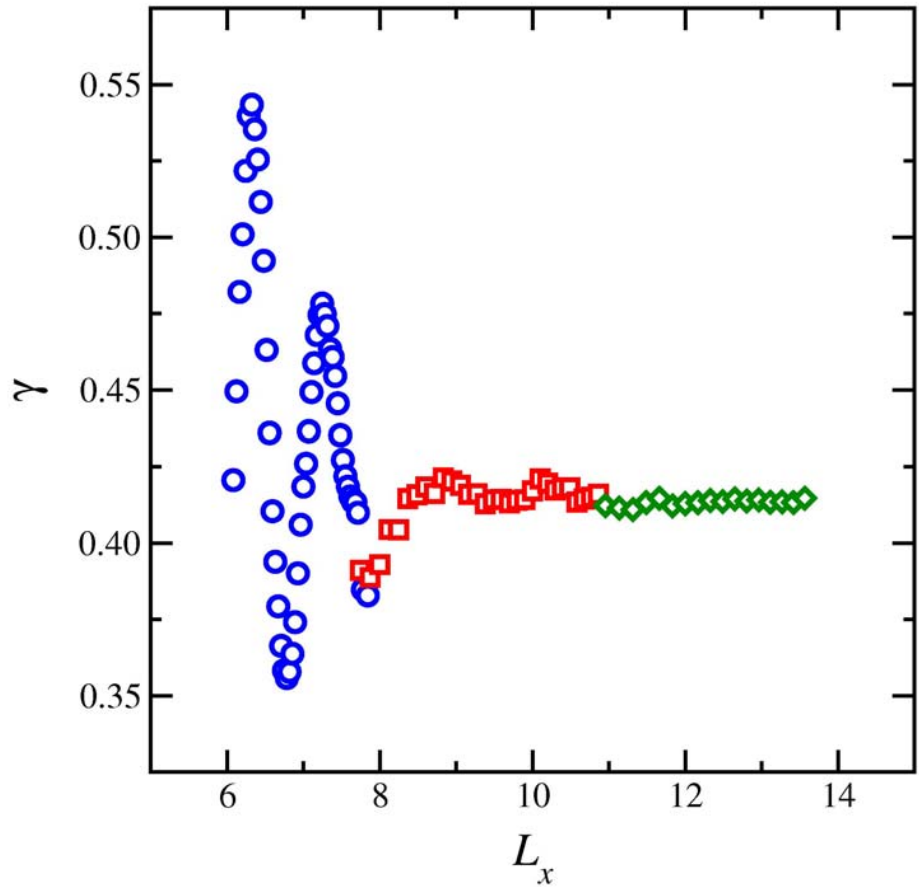

Figure 5

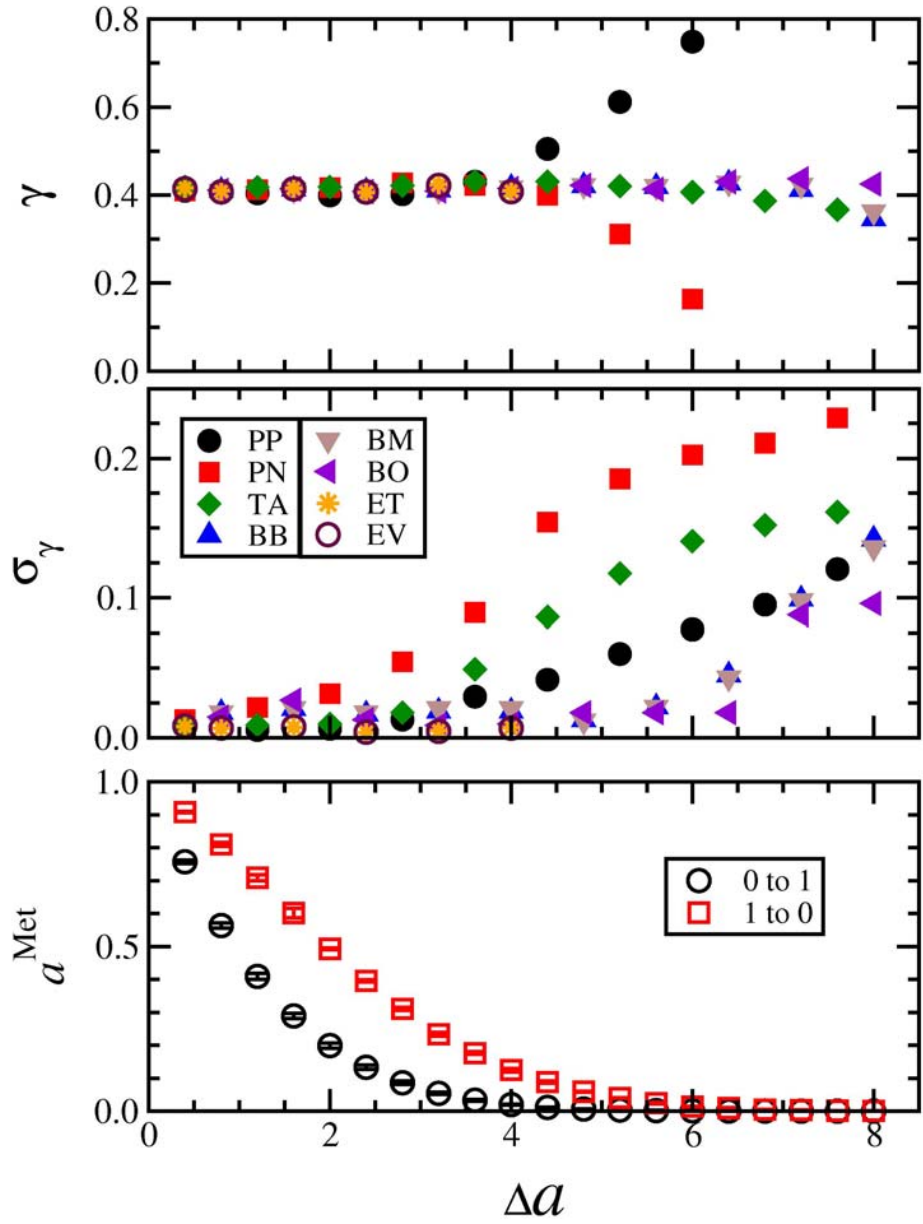

## Figure 6

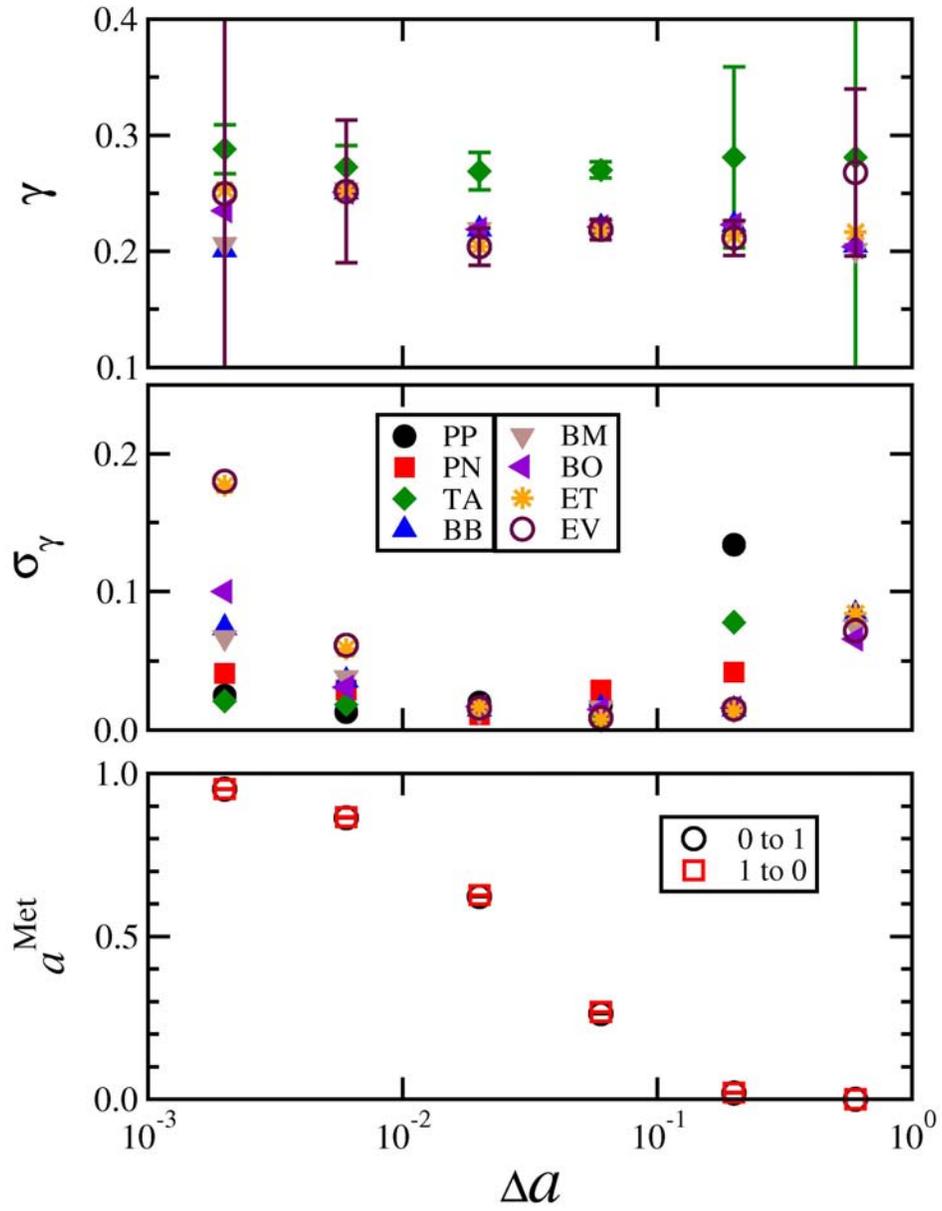

Figure 7

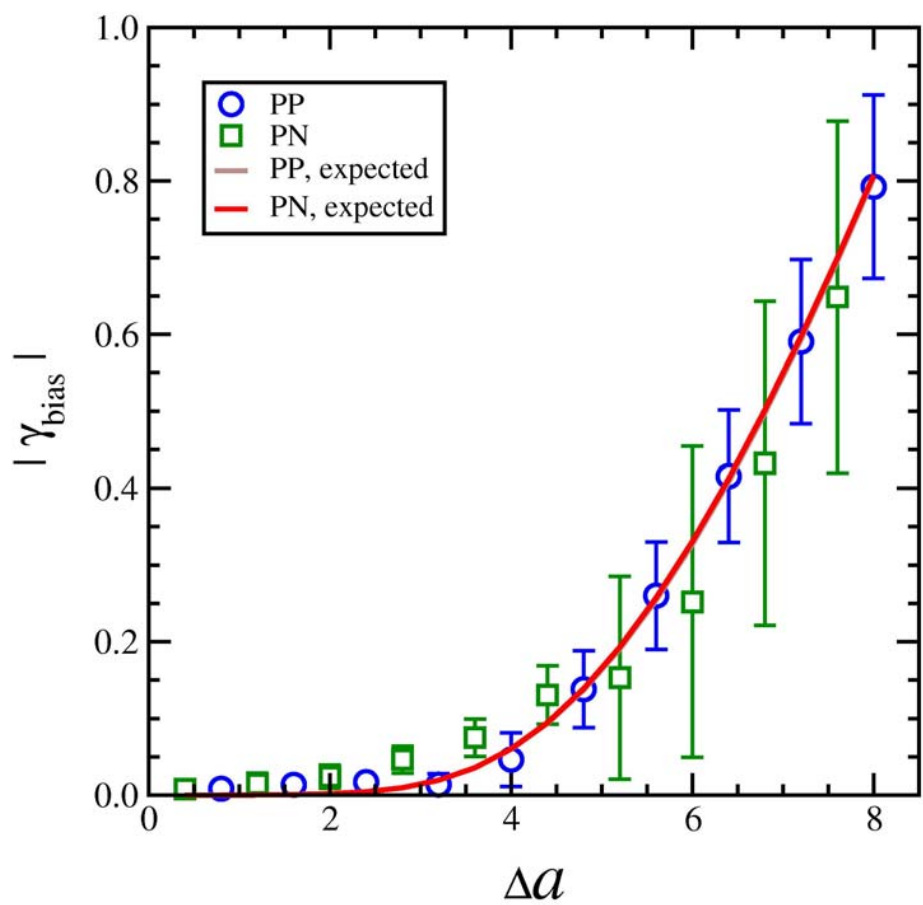